\documentclass[prd,onecolumn,preprint,nofootinbib]{revtex4}
\usepackage[plainpages=false, colorlinks=true, anchorcolor=blue, linkcolor=blue, citecolor=blue, bookmarks=false]{hyperref}

\usepackage{graphicx}
\usepackage{amsmath}
\usepackage{hyperref}
\usepackage{caption}
\usepackage{subcaption}
\usepackage{siunitx}
\usepackage{natbib}
\usepackage{placeins}
\usepackage{booktabs}
\graphicspath{{figures/}}

\begin{document}

\title{Search for high-energy neutrinos from Einstein Probe detected  Fast X-ray Transients using 14 years of IceCube muon track data}

\author{Sri Devaki Meduri}
 \altaffiliation{E-mail: ep23btech11030@iith.ac.in}
\author{Shantanu Desai}%
 \altaffiliation{E-mail: shntn05@gmail.com}

\affiliation{Department  of Physics, IIT Hyderabad,  Kandi, Telangana-502284, India}

\begin{abstract}
We perform a time-integrated search for high energy neutrinos from 27 fast X-ray transients (FXTs) detected by Einstein Probe, using 14 years of publicly available IceCube muon-track data. We employ the unbinned maximum likelihood method which is used for IceCube point source searches. We do not find any statistically significant excess from any of the FXTs. The largest TS value is obtained for EP240801a (TS=4.77), corresponding to a local $p$-value of 0.04, which is not significant. We therefore calculate 90\% confidence level upper limit on the differential muon neutrino flux.
The most stringent upper limit is obtained for EP241021a, with   $E^2 \Phi(E)=2.1\times 10^{-10}$ (GeV~cm$^{-2}$~s$^{-1}$), for a  $E^{-2}$ spectrum at a pivot energy of 100 TeV.
\end{abstract}
\maketitle
\section{Introduction}
In 2013, IceCube detected a diffuse neutrino flux observed in the TeV-PeV energy
range~\cite{IceCubescience}, whose origin is still unknown.  The first source for which a statitically significant spatial association  was established  was the blazar TXS 0506+056~\cite{IceCubeblazar}. The second such source with a  significance of $4.2\sigma$ was NGC 1068, a Seyfert galaxy~\cite{IceCubeSeyfert}. However, these sources can only account for up to  1\% of the diffuse astrophysical neutrino flux~\cite{IceCubeSeyfert,IceCube20}. 
Therefore,  in order to determine the origin  of the diffuse neutrino flux observed in IceCube, searches from a wide range of extragalactic sources have been done to look for a statistically significant excess, most of which  have reported null results~\cite{Devaki,Bouri} (and references therein).


In this work, we search for neutrino emission from fast X-ray transients (FXTs). FXTs are short duration bursts  of soft X-ray radiation that last from $10-10^4$ seconds and are located at cosmological distances~\cite{Qui,Qui24}.  Their spectral characteristics are in between classic Gamma-ray bursts (GRBs) and X-ray flashes~\cite{Angu26}. Until recently,  they were mostly identified through analysis of archival X-ray data and it was difficult to identify their host galaxies~\cite{Eappa}. The recent launch of Einstein Probe (EP) has substantially advanced the field of FXTs. The Wide-field X-ray Telescope (WXT) aboard EP observes the sky in the soft X-ray band covering 0.5-4.0 keV~\cite{EP}. 
~\cite{Guo26}.  
The Einstein Probe has reported a total of 200 FXTs since its launch two years ago. Based on an analysis of data from the first year of EP, it was shown that only about 20\% of FXTs  have gamma-ray counterparts~\cite{Zhang25,Guo26}. A number of FXTs were found to possess afterglow properties similar to that of GRBs~\cite{Li26}.  Type 1c supernovae have been detected through optical afterglow observations of many FXTs~\cite{VanDalen,Zheng26}. Type 1c (and other types of core collapse supernova) have long been associated with long GRBs~\cite{Bloom06}.
Most recently ~\citet{Connor}  has assembled a catalog of 26 spectroscopically confirmed FXTs from Einstein Probe,  and showed that their distribution is consistent with that of  long GRBs.  Therefore, it was argued that a large fraction of EP detected FXTs have a collapsar origin similar to long GRBs~\cite{Connor}.
Subsequently, it was confirmed that the formation rate and luminosity function is similar to long GRBs~\cite{Li26}. 

A  plethora of models have been invoked to explain the origin of FXTs. Possible explanations include stellar flares, supernova shock breakouts, millisecond magnetars, among others~\cite{Zheng26}. The progenitor models include  compact object mergers, tidal disruption events, proto-magnetars, etc~\cite{Yadav25}.
The weak or absent gamma-ray emission observed in most FXTs suggests that the majority of these events are unlikely to originate from standard GRB jets viewed along their axes~\cite{Zheng26}. However, as predicted in early work, off-axis and choked jets  have been shown to reproduce the observed timescales and luminosities of FXTs at the time~\cite{Rhoads2003}
Most recently, hydrodynamical simulations of a relativistic jet propagating from the progenitor to the photon diffusion radius of the cocoon have been able to explain many  of the observed properties of FXTs~\cite{Zheng26}.

Previous searches for neutrinos in coincidence with GRBs have been done by the IceCube collaboration~\cite{IceCubeGRB,IceCubeGRB17,IceCubeGRB16,IceCubeGRB15}. The main emission mechanism for production of neutrinos from GRBs is  the  Waxman-Bahcall mechanism where shock-accelerated protons in relativistic GRB jets interact with prompt gamma-ray photons via photo-meson interactions, generating charged pions that decay into high-energy neutrinos~\cite{Waxman97}. 
Neutrino production mechanisms have also been proposed from low luminosity as well as  untriggered GRBs~\cite{Murase06,LiuWang13,MuraseIoka,Tamborra15,Kashiyama13,Zhang18,Bonc19} or GRBs with choked jets~\cite{Meszaros01,Senno15}. The earliest predictions based on the  choked jet model suggested an expected detection of   0.2 neutrino-induced muons from a source at a redshift of $z \sim 1.0$ ($D_L \sim 10^{28}$ cm), assuming an isotropic equivalent energy of $10^{53}$ ergs~\cite{Meszaros01}. The diffuse 
all-flavor neutrino flux from the choked jet model has recently been estimated to  be $10^{-7} \rm{GeV~cm^{-2}~s^{-1}~sr^{-1}}$ at 30 TeV~\cite{Senno15}.

The similarity between FXTs and low-luminosity GRBs, together with the success of models invoking off-axis and choked jets to explain FXTs, motivates a search for high-energy neutrinos from these transients. This search is the primary goal of our work.
However, because  the Einstein Probe detected FXTs do not have any temporal overlap with the latest publicly available IceCube muon track data, we only carry out  time-integrated searches from FXTs using the entire catalog.  A temporal search will be deferred to future work with updated IceCube data which overlap the FXT triggers.
We also note that there are models for neutrino emission from GRBs before the gamma-ray breakout~\cite{Razzaque03,Senno15} and also in the afterglow phase~\cite{Waxman2000}. So an untriggered search for FXTs which  could pose constraints on some of these models.


This manuscript is structured as follows. The dataset used for the analysis is discussed  in Sect.~\ref{sec:dataset}. The  analysis methodology and results can be found  in Sect.~\ref{sec:analysis} and conclusions in Sect.~\ref{sec:conclusions}.

\section{Dataset}
\label{sec:dataset}
For this analysis we consider muon track data from IceCube's second data release from April 6, 2008  to  May 23, 2022~\cite{icetracks_dr2}. This data set has been referred to as IceTracks-DR2 and covers four additional years of data compared to IceTracks-DR1~\cite{icecube10yr_data}.  The IceTracks-DR2 dataset contains approximately 1.64 million neutrino events.
from four detector configurations: IC40, IC59, IC79, and IC86. All IC86 seasons have  undergone the same event
processing and use the same detector response. Further, data from IC79 to IC86-2018 have been reprocessed using updated detector calibration. The effective areas provided in this data release also incorporate improvements in detector calibration and filtering. 
A full discussion of the improvements in this catalog compared to IceTracks-DR1 can be found  in ~\cite{icetracks_dr2}.

\begin{figure}[h]
\centering
\includegraphics[width=0.7\textwidth]{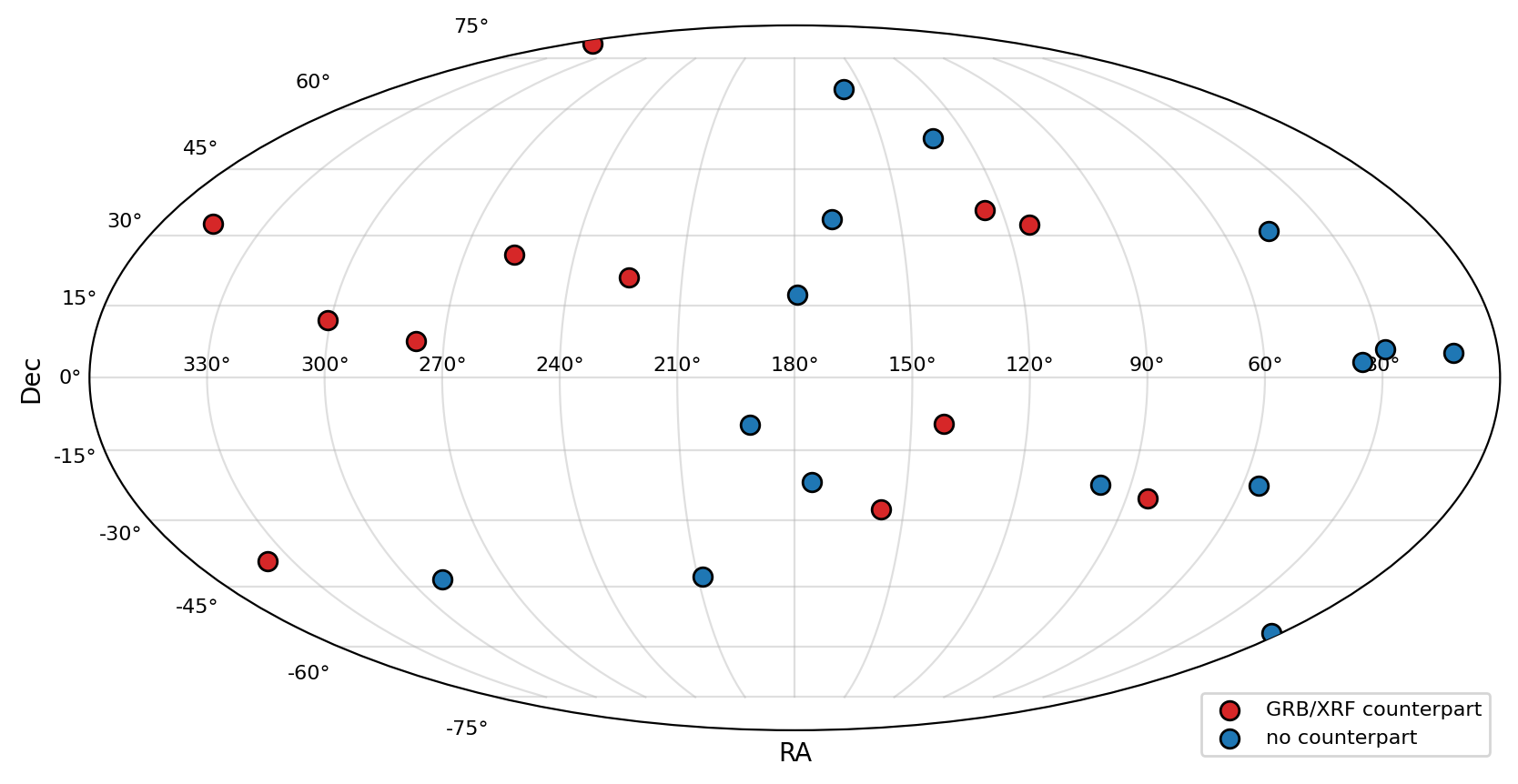}
\caption{Sky positions of the 27 EP/FXT sources. Red points have a GRB or X-ray-flash counterpart; blue points do not.}
\label{fig:skymap}
\end{figure}

The list of FXTs has mainly been obtained from the compilation in ~\cite{Connor}. This catalog was obtained by looking at the EP/WFXT transients in the energy range from 0.5-4.0 keV  until 29th August 2025,  and after culling out the flaring stars~\cite{Connor}. A total of 113 such transients were obtained from the publicly available data. Among these, 26 FXTs had  spectroscopic redshifts resulting in a redshift completeness of $\sim$ 23\%. The spectroscopic redshifts were obtained from public GCN circulars or  through dedicated spectroscopic observations with VLT for some FXTs such as EP250302a, EP250821a, and EP250827a!\cite{Connor}.
Twelve of these FXTs have a GRB counterparts, whereas no GRB counterparts were detected for the remaining sources.
This catalog has been used to estimate  the  luminosity function and formation rate of FXTs~\cite{Guo26,Li26}.
 We also added the recently detected FXT, viz. EP260119a located at a redshift of $z=5.47$~\cite{Angu26}.  A skymap of the 27 FXTs used for this analysis can be found in Fig.~\ref{fig:skymap}. 


\section{Analysis and Results}
\label{sec:analysis}
For this work, we use the unbinned maximum likelihood method~\cite{Montaruli} which is used in IceCube point source analysis~\cite{IceCube20}. This analysis was performed  using  the  {\tt SkyLLH}~\cite{skyllh_extension,skyllh2,skyllh21} software package.  We describe the procedure below.

For a dataset containing $N$ events, of which $n_s$ are assumed to originate from the source, the likelihood function can be defined as follows:
\begin{equation}
    \mathcal{L} (n_s,\gamma)  = \prod_{i=1}^N \left(\frac{n_s}{N} S_i (\gamma) + \left(1 - \frac{n_s}{N}\right) B_i\right),
\label{eq:1}
\end{equation}
where $\gamma$ is the spectral index, $B_i$ and $S_i$ denote the background and signal PDF respectively for $i$-th neutrino event. The signal pdf depends on  both a spatial and an energy dependent term, of which only the energy dependent term depends on $\gamma$. The background PDF can be extracted  from the data. More details about the models used for $B_i$ and $S_i$ can be found in ~\cite{skyllh2} and references therein. We then construct the test statistics defined as follows:
\begin{equation}
    \rm{TS} = 2 \log \frac{\mathcal{L}(\hat{n}_s,\hat{\gamma})}{\mathcal{L}(n_s=0)},
\label{eq:ts}
\end{equation}
where  $\hat{n}_s$ and $\hat{\gamma}$  are the values of $n_s$ and $\gamma$, which  maximize the likelihood in Eq.~\ref{eq:ts}. For a large number of signal events, TS asymptotically follows a  $\chi^2$ distribution for two degrees of freedom~\cite{Wilks}. However, because $n_s$ and $\gamma$ are correlated  the distribution of the null hypothesis could lead to deviations from $\chi^2$ distribution for two degrees of freedom~\cite{Luo26}. 
For this purpose we use background trials  using the observed data to calculate the $p$-value similar to recent works which have used the same software for point-source searches~\cite{Chen26}. The distribution from background trials is obtained using the same  {\tt SkyLLH} software.
For a statistically significant detection  of $5\sigma$, the local  $p$-value for each source has to be smaller than  $3.0 \times 10^{-7}$~\cite{Cowan11,Ganguly17}.  Accounting for the look-elsewhere effect associated with the  27 FXTs, the local $p$-value should be even smaller if the global $p$-value should exceed $5\sigma$.

For every source the unbinned likelihood
ratio is then maximized over the number of signal events $n_s\ge 0$ and the spectral index $\gamma$ of a power-law spectrum 
giving a per-source test statistic $\mathrm{TS}$. 
The best-fit values can be found in Table~\ref{tab:ep_catalog}. Fifteen of the 27 positions return a non-zero TS; the other twelve
converge to $\mathrm{TS}=0$, $\hat{n}_s=0$. The largest values are EP240801a
($\mathrm{TS}=4.77$, $\hat{n}_s\simeq 29$, $\gamma\simeq 3.3$), EP250704a
(TS=3.74), EP241217b (TS=2.90), EP250821a (TS=2.76) and EP240315a (TS=2.59); all
other sources have $\mathrm{TS}<2.3$. The source EP240801a, which has  the largest TS value (TS=4.77) has a local $p$-value of 0.04. 
Hence, the largest TS value is not significant enough  to claim a statistically significant detection.

We therefore derive flux upper limits for each  FXT. For this purpose, we carry out trials to obtain the upper limit on neutrino flux by injecting varying  number of signal neutrinos~\cite{Chen26,WangUniv}. We then calculate the $p$-value for obtaining a value larger than the observed $\hat{n}_s$ for the real data. The 90\%  confidence level (c.l.) upper limit on the number of signal events $n_s^{90}$ is then given by  the value of injected signal for which $p$-value is equal to 0.1.   Our limits on $n_s^{90}$ (cf. Table~\ref{tab:ep_catalog}) are roughly an order of magnitude larger than the predicted muon neutrino rate from choked jet model~\cite{Meszaros01} for $E_{iso} \sim 10^{53}$ ergs and $z=1.0$. However, since  $E_{iso}$ for most FXTs is less than $10^{53}$ ergs, we  cannot rule out this model.  However, we would expect more robust constraints from a triggered search.

To convert this limit into a differential neutrino flux limit, we assume a power-law form for the neutrino spectrum:
\begin{equation}
    \Phi(E) = \Phi_0 \left( \frac{E}{E_0} \right)^{-\gamma},
    \label{eq:phinu}
\end{equation}
where $\Phi_0$ is the normalization at $E_0=1000$ TeV and $\Gamma$ is the spectral index, for which we assume $\Gamma=2.0$ for all sources, independent of the best-fit $\gamma$ found for each source.  The differential 
muon neutrino flux limit ($E^2\Phi (E)$) is then calculated by dividing $n_s^{90}$  by exposure and effective area, as outlined in one of our previous works~\cite{Devaki}.
These limits can also  be found in Table~\ref{tab:ep_catalog} and have values in the range of  $\mathcal{O} (10^{-10})$ GeV$\rm{cm}^{-2}~\rm{s}^{-1}$. The most stringent limit is obtained for EP241021a, which is  given by  $E^2 \Phi (E) =2.1 \times 10^{-10}$ GeV~cm$^{-2}$~s$^{-1}$ at a pivot energy of 100 TeV.



\begin{table}[h]
\centering
\scriptsize
\setlength{\tabcolsep}{4pt}
\caption{Einstein Probe fast X-ray transient (EP/FXT) sample analyzed as steady-state point sources, with the best-fit test statistic, number of signal events and spectral index from the standard single-source SkyLLH analysis (time-integrated). The spectral index is bounded to $1\le\gamma\le4$ in the fit and is unconstrained for sources with $n_s=0$. $p_{\rm local}$ is the fraction of background-only trials for that source with a TS at least as large as observed. The second last column ($n_s^{90}$) is the 90\% c.l. upper limit on the number of signal events. The last column is the 90\% CL upper limit on differential muon neutrino flux $E^2 \Phi(E)$ at 100~TeV (GeV~cm$^{-2}$~s$^{-1}$, assuming an  $E^{-2}$ spectrum).}
\label{tab:ep_catalog}
\begin{tabular}{|l|c|c|c|c|c|c|c|c|c|}
\toprule\hline
Source & $z$ & RA & Dec & TS & $\hat{n}_s$ & $\hat{\gamma}$ & $p_{\rm local}$ & $n_s^{90}$ & $E^2 \Phi(E)$ (90\% CL) \\
 & & (deg) & (deg) & & & & & & (GeVcm$^{-2}$s$^{-1}$) \\
\hline
\midrule
EP240315a & 4.859 & 141.644 & $-$9.547 & 2.59 & 7.6 & 2.90 & 0.053 & 5.4 & $7.4\times10^{-10}$ \\
EP240414a & 0.401 & 191.498 & $-$9.695 & 0.00 & 0.0 & 2.90 & 1.000 & 3.3 & $4.6\times10^{-10}$ \\
EP240801a & 1.673 & 345.140 & 32.610 & 4.77 & 28.8 & 3.29 & 0.040 & 19 & $8.3\times10^{-10}$ \\
EP240804a & 3.662 & 337.644 & $-$39.121 & 0.00 & 0.0 & 1.75 & 1.000 & 5.3 & $2.9\times10^{-9}$  \\
EP240806a & 2.818 & 11.491 & 5.091 & 0.55 & 12.5 & 3.66 & 0.404 & 7 & $2.2\times10^{-10}$ \\
EP241021a & 0.748 & 28.852 & 5.957 & 0.00 & 0.0 & 4.00 & 1.000 & 6.3 & $2.1\times10^{-10}$  \\
EP241030a & 1.411 & 343.013 & 80.449 & 0.00 & 0.0 & 4.00 & 1.000 & 14.8 & $8.0\times10^{-10}$  \\
EP241107a & 0.456 & 35.008 & 3.333 & 2.21 & 23.4 & 3.29 & 0.160 & 10.4 & $3.2\times10^{-10}$ \\
EP241113a & 1.53 & 131.996 & 52.382 & 0.00 & 0.0 & 4.00 & 1.000 & 7.1 & $3.5\times10^{-10}$ \\
EP241217a & 4.59 & 46.957 & 30.901 & 0.00 & 0.0 & 4.00 & 1.000 & 7 & $3.0\times10^{-10}$\\
EP241217b & 1.879 & 84.167 & $-$25.281 & 2.90 & 8.7 & 3.04 & 0.094 & 8.3 & $3.1\times10^{-9}$ \\
EP250108a & 0.176 & 55.623 & $-$22.509 & 0.07 & 1.0 & 4.00 & 0.607 & 3.6 & $1.2\times10^{-9}$  \\
EP250125a & 2.89 & 175.364 & $-$21.708 & 0.00 & 0.0 & 1.62 & 1.000 & 3.7 & $1.2\times10^{-9}$ \\
EP250205a & 3.55 & 113.522 & 32.363 & 0.00 & 0.0 & 4.00 & 1.000 & 7.1 & $3.1\times10^{-10}$  \\
EP250215a & 4.61 & 156.343 & $-$27.704 & 1.55 & 6.8 & 3.25 & 0.176 & 7.1 & $2.9\times10^{-9}$ \\
EP250223a & 2.756 & 98.258 & $-$22.432 & 0.00 & 0.0 & 4.00 & 1.000 & 4.1 & $1.4\times10^{-9}$  \\
EP250226a & 3.315 & 224.273 & 20.973 & 0.01 & 1.2 & 3.29 & 0.631 & 8 & $3.1\times10^{-10}$  \\
EP250302a & 1.131 & 169.515 & 33.585 & 0.61 & 11.6 & 3.48 & 0.352 & 10.3 & $4.6\times10^{-10}$ \\
EP250304a & 0.2 & 208.394 & $-$42.818 & 1.18 & 2.3 & 4.00 & 0.197 & 9 & $5.3\times10^{-9}$ \\
EP250321a & 4.368 & 179.262 & 17.386 & 0.35 & 8.9 & 3.71 & 0.394 & 7.6 & $2.9\times10^{-10}$ \\
EP250404a & 1.88 & 125.064 & 35.511 & 0.00 & 0.0 & 4.00 & 1.000 & 7.2 & $3.2\times10^{-10}$  \\
EP250416a & 0.963 & 256.402 & 25.768 & 0.00 & 0.0 & 4.00 & 1.000 & 7.2 & $3.0\times10^{-10}$  \\
EP250427a & 1.52 & 277.281 & 7.570 & 1.69 & 19.0 & 4.00 & 0.162 & 11.8 & $3.9\times10^{-10}$ \\
EP250704a & 0.661 & 300.872 & 12.030 & 3.74 & 30.8 & 3.86 & 0.052 & 13.8 & $4.9\times10^{-10}$ \\
EP250821a & 0.577 & 289.856 & $-$43.410 & 2.76 & 6.6 & 3.95 & 0.079 & 10.1 & $6.1\times10^{-9}$ \\
EP250827a & 1.61 & 3.488 & $-$56.463 & 0.00 & 0.0 & 3.26 & 1.000 & 5.7 & $4.0\times10^{-9}$ \\
EP260119a & 5.47 & 158.105 & 65.498 & 1.24 & 16.3 & 3.54 & 0.157 & 20.3 & $1.1\times10^{-9}$ \\
\hline
\bottomrule
\end{tabular}
\end{table}

\section{Conclusions}
\label{sec:conclusions}
In this work, we search for steady-state high energy neutrino emission from 27 spectroscopically confirmed FXTs detected between 2024 and 2026 by the Einstein Probe X-ray satellite using the recently released 14 year muon track data from IceCube from 2008-2022. For this purpose, we use the standard unbinned maximum likelihood analysis which has been used for IceCube point source search using the {\tt SkyLLH} software package.   Our results are summarized in Table~\ref{tab:ep_catalog}. We do not find any statistically significant excess for any FXT. The maximum TS value is observed for EP240801a, whose local $p$-value is 0.04 and hence is not significant even without considering the look-elsewhere effect. We then calculate 90\% c.l. differential muon neutrino flux limit at 100 TeV for each of the 27 FXTs. The most stringent upper limit is observed for EP241021a, given by   $E^2\,dF/dE=2.1 \times 10^{-10}$ GeV~cm$^{-2}$~s$^{-1}$, for a  $E^{-2}$ spectrum.
Our limits are about an order of magnitude compared to some of the early predictions from the choked GRB jet model~\cite{Meszaros01}.
In a future work, we will supplement this analysis with a temporal analysis using narrow time windows around the time of the detected FXTs to further increase the sensitivity, once the updated IceCube point source catalog is released. 

\section*{Acknowledgements}
The computational work used for this analysis was supported by the National Supercomputing Mission (NSM), Government of India, through access to the ``PARAM SEVA'' facility at IIT Hyderabad. The NSM is implemented by the Centre for Development of Advanced Computing (C-DAC) with funding from the Ministry of Electronics and Information Technology (MeitY) and the Department of Science and Technology (DST).
We have used OpenAI chatgpt to correct for language and stylistic issues in the manuscript.
\bibliography{ref}

@article{icecube10yr_data,
  author  = {{IceCube Collaboration} and Abbasi, R. and others},
  title   = {IceCube Data for Neutrino Point-Source Searches Years 2008--2018},
  journal = {arXiv e-prints},
  year    = {2021},
  eprint  = {2101.09836},
  archivePrefix = {arXiv},
  primaryClass  = {astro-ph.HE}
}

@article{IceCubescience,
    author = "Aartsen, M. G. and others",
    collaboration = "IceCube",
    title = "{Evidence for High-Energy Extraterrestrial Neutrinos at the IceCube Detector}",
    eprint = "1311.5238",
    archivePrefix = "arXiv",
    primaryClass = "astro-ph.HE",
    doi = "10.1126/science.1242856",
    journal = "Science",
    volume = "342",
    pages = "1242856",
    year = "2013"
}

@article{IceCubeblazar,
    author = "Aartsen, M. G. and others",
    collaboration = "IceCube, Fermi-LAT, MAGIC, AGILE, ASAS-SN, HAWC, H.E.S.S., INTEGRAL, Kanata, Kiso, Kapteyn, Liverpool Telescope, Subaru, Swift NuSTAR, VERITAS, VLA/17B-403",
    title = "{Multimessenger observations of a flaring blazar coincident with high-energy neutrino IceCube-170922A}",
    eprint = "1807.08816",
    archivePrefix = "arXiv",
    primaryClass = "astro-ph.HE",
    doi = "10.1126/science.aat1378",
    journal = "Science",
    volume = "361",
    number = "6398",
    pages = "eaat1378",
    year = "2018"
}

@ARTICLE{IceCubeGRB,
       author = {{Abbasi}, R. and {Ackermann}, M. and {Adams}, J. and {Aguilar}, J.~A. and {Ahlers}, M. and {Ahrens}, M. and {Alameddine}, J.~M. and {Alves}, A.~A. and {Amin}, N.~M. and {Andeen}, K. and {Anderson}, T. and {Anton}, G. and {Arg{\"u}elles}, C. and {Ashida}, Y. and {Athanasiadou}, S. and {Axani}, S. and {Bai}, X. and {Balagopal}, A.~V. and {Barwick}, S.~W. and {Basu}, V. and {Baur}, S. and {Bay}, R. and {Beatty}, J.~J. and {Becker}, K.-H. and {Becker Tjus}, J. and {Beise}, J. and {Bellenghi}, C. and {Benda}, S. and {BenZvi}, S. and {Berley}, D. and {Bernardini}, E. and {Besson}, D.~Z. and {Binder}, G. and {Bindig}, D. and {Blaufuss}, E. and {Blot}, S. and {Boddenberg}, M. and {Bontempo}, F. and {Book}, J.~Y. and {Borowka}, J. and {B{\"o}ser}, S. and {Botner}, O. and {B{\"o}ttcher}, J. and {Bourbeau}, E. and {Bradascio}, F. and {Braun}, J. and {Brinson}, B. and {Bron}, S. and {Brostean-Kaiser}, J. and {Burley}, R.~T. and {Busse}, R.~S. and {Campana}, M.~A. and {Carnie-Bronca}, E.~G. and {Chen}, C. and {Chen}, Z. and {Chirkin}, D. and {Choi}, K. and {Clark}, B.~A. and {Clark}, K. and {Classen}, L. and {Coleman}, A. and {Collin}, G.~H. and {Connolly}, A. and {Conrad}, J.~M. and {Coppin}, P. and {Correa}, P. and {Cowen}, D.~F. and {Cross}, R. and {Dappen}, C. and {Dave}, P. and {De Clercq}, C. and {DeLaunay}, J.~J. and {L{\'o}pez}, D. Delgado and {Dembinski}, H. and {Deoskar}, K. and {Desai}, A. and {Desiati}, P. and {de Vries}, K.~D. and {de Wasseige}, G. and {DeYoung}, T. and {Diaz}, A. and {D{\'\i}az-V{\'e}lez}, J.~C. and {Dittmer}, M. and {Dujmovic}, H. and {DuVernois}, M.~A. and {Ehrhardt}, T. and {Eller}, P. and {Engel}, R. and {Erpenbeck}, H. and {Evans}, J. and {Evenson}, P.~A. and {Fan}, K.~L. and {Fazely}, A.~R. and {Fedynitch}, A. and {Feigl}, N. and {Fiedlschuster}, S. and {Fienberg}, A.~T. and {Finley}, C. and {Fischer}, L. and {Fox}, D. and {Franckowiak}, A. and {Friedman}, E. and {Fritz}, A. and {F{\"u}rst}, P. and {Gaisser}, T.~K. and {Gallagher}, J. and {Ganster}, E. and {Garcia}, A. and {Garrappa}, S. and {Gerhardt}, L. and {Ghadimi}, A. and {Glaser}, C. and {Glauch}, T. and {Gl{\"u}senkamp}, T. and {Goehlke}, N. and {Gonzalez}, J.~G. and {Goswami}, S. and {Grant}, D. and {Gr{\'e}goire}, T. and {Griswold}, S. and {G{\"u}nther}, C. and {Gutjahr}, P. and {Haack}, C. and {Hallgren}, A. and {Halliday}, R. and {Halve}, L. and {Halzen}, F. and {Ha Minh}, M. and {Hanson}, K. and {Hardin}, J. and {Harnisch}, A.~A. and {Haungs}, A. and {Helbing}, K. and {Henningsen}, F. and {Hettinger}, E.~C. and {Hickford}, S. and {Hignight}, J. and {Hill}, C. and {Hill}, G.~C. and {Hoffman}, K.~D. and {Hoshina}, K. and {Hou}, W. and {Huber}, M. and {Huber}, T. and {Hultqvist}, K. and {H{\"u}nnefeld}, M. and {Hussain}, R. and {Hymon}, K. and {In}, S. and {Iovine}, N. and {Ishihara}, A. and {Jansson}, M. and {Japaridze}, G.~S. and {Jeong}, M. and {Jin}, M. and {Jones}, B.~J.~P. and {Kang}, D. and {Kang}, W. and {Kang}, X. and {Kappes}, A. and {Kappesser}, D. and {Kardum}, L. and {Karg}, T. and {Karl}, M. and {Karle}, A. and {Katz}, U. and {Kauer}, M. and {Kellermann}, M. and {Kelley}, J.~L. and {Kheirandish}, A. and {Kin}, K. and {Kiryluk}, J. and {Klein}, S.~R. and {Kochocki}, A. and {Koirala}, R. and {Kolanoski}, H. and {Kontrimas}, T. and {K{\"o}pke}, L. and {Kopper}, C. and {Kopper}, S. and {Koskinen}, D.~J. and {Koundal}, P. and {Kovacevich}, M. and {Kowalski}, M. and {Kozynets}, T. and {Krupczak}, E. and {Kun}, E. and {Kurahashi}, N. and {Lad}, N. and {Lagunas Gualda}, C. and {Larson}, M.~J. and {Lauber}, F. and {Lazar}, J.~P. and {Lee}, J.~W. and {Leonard}, K. and {Leszczy{\'n}ska}, A. and {Li}, Y. and {Lincetto}, M. and {Liu}, Q.~R. and {Liubarska}, M.},
        title = "{Searches for Neutrinos from Gamma-Ray Bursts Using the IceCube Neutrino Observatory}",
      journal = {\apj},
         year = 2022,
        month = nov,
       volume = {939},
       number = {2},
          eid = {116},
        pages = {116},
          doi = {10.3847/1538-4357/ac9785},
archivePrefix = {arXiv},
       eprint = {2205.11410},
 primaryClass = {astro-ph.HE},
       adsurl = {https://ui.adsabs.harvard.edu/abs/2022ApJ...939..116A}
}

@ARTICLE{IceCubeGRB17,
       author = {{Aartsen}, M.~G. and {Ackermann}, M. and {Adams}, J. and {Aguilar}, J.~A. and {Ahlers}, M. and {Ahrens}, M. and {Samarai}, I. Al and {Altmann}, D. and {Andeen}, K. and {Anderson}, T. and {Ansseau}, I. and {Anton}, G. and {Archinger}, M. and {Arg{\"u}elles}, C. and {Auffenberg}, J. and {Axani}, S. and {Bai}, X. and {Barwick}, S.~W. and {Baum}, V. and {Bay}, R. and {Beatty}, J.~J. and {Becker Tjus}, J. and {Becker}, K.-H. and {BenZvi}, S. and {Berley}, D. and {Bernardini}, E. and {Besson}, D.~Z. and {Binder}, G. and {Bindig}, D. and {Blaufuss}, E. and {Blot}, S. and {Bohm}, C. and {B{\"o}rner}, M. and {Bos}, F. and {Bose}, D. and {B{\"o}ser}, S. and {Botner}, O. and {Braun}, J. and {Brayeur}, L. and {Bretz}, H.-P. and {Bron}, S. and {Burgman}, A. and {Carver}, T. and {Casier}, M. and {Cheung}, E. and {Chirkin}, D. and {Christov}, A. and {Clark}, K. and {Classen}, L. and {Coenders}, S. and {Collin}, G.~H. and {Conrad}, J.~M. and {Cowen}, D.~F. and {Cross}, R. and {Day}, M. and {de Andr{\'e}}, J.~P.~A.~M. and {De Clercq}, C. and {del Pino Rosendo}, E. and {Dembinski}, H. and {De Ridder}, S. and {Desiati}, P. and {de Vries}, K.~D. and {de Wasseige}, G. and {de With}, M. and {DeYoung}, T. and {D{\'\i}az-V{\'e}lez}, J.~C. and {di Lorenzo}, V. and {Dujmovic}, H. and {Dumm}, J.~P. and {Dunkman}, M. and {Eberhardt}, B. and {Ehrhardt}, T. and {Eichmann}, B. and {Eller}, P. and {Euler}, S. and {Evenson}, P.~A. and {Fahey}, S. and {Fazely}, A.~R. and {Feintzeig}, J. and {Felde}, J. and {Filimonov}, K. and {Finley}, C. and {Flis}, S. and {F{\"o}sig}, C.-C. and {Franckowiak}, A. and {Friedman}, E. and {Fuchs}, T. and {Gaisser}, T.~K. and {Gallagher}, J. and {Gerhardt}, L. and {Ghorbani}, K. and {Giang}, W. and {Gladstone}, L. and {Glauch}, T. and {Gl{\"u}senkamp}, T. and {Goldschmidt}, A. and {Gonzalez}, J.~G. and {Grant}, D. and {Griffith}, Z. and {Haack}, C. and {Hallgren}, A. and {Halzen}, F. and {Hansen}, E. and {Hansmann}, T. and {Hanson}, K. and {Hebecker}, D. and {Heereman}, D. and {Helbing}, K. and {Hellauer}, R. and {Hickford}, S. and {Hignight}, J. and {Hill}, G.~C. and {Hoffman}, K.~D. and {Hoffmann}, R. and {Hoshina}, K. and {Huang}, F. and {Huber}, M. and {Hultqvist}, K. and {In}, S. and {Ishihara}, A. and {Jacobi}, E. and {Japaridze}, G.~S. and {Jeong}, M. and {Jero}, K. and {Jones}, B.~J.~P. and {Kang}, W. and {Kappes}, A. and {Karg}, T. and {Karle}, A. and {Katz}, U. and {Kauer}, M. and {Keivani}, A. and {Kelley}, J.~L. and {Kheirandish}, A. and {Kim}, J. and {Kim}, M. and {Kintscher}, T. and {Kiryluk}, J. and {Kittler}, T. and {Klein}, S.~R. and {Kohnen}, G. and {Koirala}, R. and {Kolanoski}, H. and {Konietz}, R. and {K{\"o}pke}, L. and {Kopper}, C. and {Kopper}, S. and {Koskinen}, D.~J. and {Kowalski}, M. and {Krings}, K. and {Kroll}, M. and {Kr{\"u}ckl}, G. and {Kr{\"u}ger}, C. and {Kunnen}, J. and {Kunwar}, S. and {Kurahashi}, N. and {Kuwabara}, T. and {Kyriacou}, A. and {Labare}, M. and {Lanfranchi}, J.~L. and {Larson}, M.~J. and {Lauber}, F. and {Lennarz}, D. and {Lesiak-Bzdak}, M. and {Leuermann}, M. and {Lu}, L. and {L{\"u}nemann}, J. and {Madsen}, J. and {Maggi}, G. and {Mahn}, K.~B.~M. and {Mancina}, S. and {Mandelartz}, M. and {Maruyama}, R. and {Mase}, K. and {Maunu}, R. and {McNally}, F. and {Meagher}, K. and {Medici}, M. and {Meier}, M. and {Menne}, T. and {Merino}, G. and {Meures}, T. and {Miarecki}, S. and {Micallef}, J. and {Moment{\'e}}, G. and {Montaruli}, T. and {Moulai}, M. and {Nahnhauer}, R. and {Naumann}, U. and {Neer}, G. and {Niederhausen}, H. and {Nowicki}, S.~C. and {Nygren}, D.~R. and {Obertacke Pollmann}, A. and {Olivas}, A. and {O'Murchadha}, A. and {Palczewski}, T. and {Pandya}, H. and {Pankova}, D.~V. and {Peiffer}, P.},
        title = "{Extending the Search for Muon Neutrinos Coincident with Gamma-Ray Bursts in IceCube Data}",
      journal = {\apj},
         year = 2017,
        month = jul,
       volume = {843},
       number = {2},
          eid = {112},
        pages = {112},
          doi = {10.3847/1538-4357/aa7569},
archivePrefix = {arXiv},
       eprint = {1702.06868},
 primaryClass = {astro-ph.HE},
       adsurl = {https://ui.adsabs.harvard.edu/abs/2017ApJ...843..112A}
}

@ARTICLE{IceCubeGRB16,
       author = {{Aartsen}, M.~G. and {Abraham}, K. and {Ackermann}, M. and {Adams}, J. and {Aguilar}, J.~A. and {Ahlers}, M. and {Ahrens}, M. and {Altmann}, D. and {Anderson}, T. and {Ansseau}, I. and {Anton}, G. and {Archinger}, M. and {Arguelles}, C. and {Arlen}, T.~C. and {Auffenberg}, J. and {Bai}, X. and {Barwick}, S.~W. and {Baum}, V. and {Bay}, R. and {Beatty}, J.~J. and {Becker Tjus}, J. and {Becker}, K.-H. and {Beiser}, E. and {BenZvi}, S. and {Berghaus}, P. and {Berley}, D. and {Bernardini}, E. and {Bernhard}, A. and {Besson}, D.~Z. and {Binder}, G. and {Bindig}, D. and {Bissok}, M. and {Blaufuss}, E. and {Blumenthal}, J. and {Boersma}, D.~J. and {Bohm}, C. and {B{\"o}rner}, M. and {Bos}, F. and {Bose}, D. and {B{\"o}ser}, S. and {Botner}, O. and {Braun}, J. and {Brayeur}, L. and {Bretz}, H.-P. and {Buzinsky}, N. and {Casey}, J. and {Casier}, M. and {Cheung}, E. and {Chirkin}, D. and {Christov}, A. and {Clark}, K. and {Classen}, L. and {Coenders}, S. and {Collin}, G.~H. and {Conrad}, J.~M. and {Cowen}, D.~F. and {Cruz Silva}, A.~H. and {Daughhetee}, J. and {Davis}, J.~C. and {Day}, M. and {de Andr{\'e}}, J.~P.~A.~M. and {De Clercq}, C. and {del Pino Rosendo}, E. and {Dembinski}, H. and {De Ridder}, S. and {Desiati}, P. and {de Vries}, K.~D. and {de Wasseige}, G. and {de With}, M. and {DeYoung}, T. and {D{\'\i}az-V{\'e}lez}, J.~C. and {di Lorenzo}, V. and {Dujmovic}, H. and {Dumm}, J.~P. and {Dunkman}, M. and {Eberhardt}, B. and {Ehrhardt}, T. and {Eichmann}, B. and {Euler}, S. and {Evenson}, P.~A. and {Fahey}, S. and {Fazely}, A.~R. and {Feintzeig}, J. and {Felde}, J. and {Filimonov}, K. and {Finley}, C. and {Flis}, S. and {F{\"o}sig}, C.-C. and {Fuchs}, T. and {Gaisser}, T.~K. and {Gaior}, R. and {Gallagher}, J. and {Gerhardt}, L. and {Ghorbani}, K. and {Gier}, D. and {Gladstone}, L. and {Glagla}, M. and {Gl{\"u}senkamp}, T. and {Goldschmidt}, A. and {Golup}, G. and {Gonzalez}, J.~G. and {G{\'o}ra}, D. and {Grant}, D. and {Griffith}, Z. and {Ha}, C. and {Haack}, C. and {Haj Ismail}, A. and {Hallgren}, A. and {Halzen}, F. and {Hansen}, E. and {Hansmann}, B. and {Hansmann}, T. and {Hanson}, K. and {Hebecker}, D. and {Heereman}, D. and {Helbing}, K. and {Hellauer}, R. and {Hickford}, S. and {Hignight}, J. and {Hill}, G.~C. and {Hoffman}, K.~D. and {Hoffmann}, R. and {Holzapfel}, K. and {Homeier}, A. and {Hoshina}, K. and {Huang}, F. and {Huber}, M. and {Huelsnitz}, W. and {Hulth}, P.~O. and {Hultqvist}, K. and {In}, S. and {Ishihara}, A. and {Jacobi}, E. and {Japaridze}, G.~S. and {Jeong}, M. and {Jero}, K. and {Jones}, B.~J.~P. and {Jurkovic}, M. and {Kappes}, A. and {Karg}, T. and {Karle}, A. and {Katz}, U. and {Kauer}, M. and {Keivani}, A. and {Kelley}, J.~L. and {Kemp}, J. and {Kheirandish}, A. and {Kim}, M. and {Kintscher}, T. and {Kiryluk}, J. and {Klein}, S.~R. and {Kohnen}, G. and {Koirala}, R. and {Kolanoski}, H. and {Konietz}, R. and {K{\"o}pke}, L. and {Kopper}, C. and {Kopper}, S. and {Koskinen}, D.~J. and {Kowalski}, M. and {Krings}, K. and {Kroll}, G. and {Kroll}, M. and {Kr{\"u}ckl}, G. and {Kunnen}, J. and {Kunwar}, S. and {Kurahashi}, N. and {Kuwabara}, T. and {Labare}, M. and {Lanfranchi}, J.~L. and {Larson}, M.~J. and {Lennarz}, D. and {Lesiak-Bzdak}, M. and {Leuermann}, M. and {Leuner}, J. and {Lu}, L. and {L{\"u}nemann}, J. and {Madsen}, J. and {Maggi}, G. and {Mahn}, K.~B.~M. and {Mandelartz}, M. and {Maruyama}, R. and {Mase}, K. and {Matis}, H.~S. and {Maunu}, R. and {McNally}, F. and {Meagher}, K. and {Medici}, M. and {Meier}, M. and {Meli}, A. and {Menne}, T. and {Merino}, G. and {Meures}, T. and {Miarecki}, S. and {Middell}, E. and {Mohrmann}, L. and {Montaruli}, T. and {Morse}, R. and {Nahnhauer}, R. and {Naumann}, U.},
        title = "{An All-sky Search for Three Flavors of Neutrinos from Gamma-ray Bursts with the IceCube Neutrino Observatory}",
      journal = {\apj},
         year = 2016,
        month = jun,
       volume = {824},
       number = {2},
          eid = {115},
        pages = {115},
          doi = {10.3847/0004-637X/824/2/115},
archivePrefix = {arXiv},
       eprint = {1601.06484},
 primaryClass = {astro-ph.HE},
       adsurl = {https://ui.adsabs.harvard.edu/abs/2016ApJ...824..115A}
}

@ARTICLE{Waxman97,
       author = {{Waxman}, Eli and {Bahcall}, John},
        title = "{High Energy Neutrinos from Cosmological Gamma-Ray Burst Fireballs}",
      journal = {\prl},
         year = 1997,
        month = mar,
       volume = {78},
       number = {12},
        pages = {2292-2295},
          doi = {10.1103/PhysRevLett.78.2292},
archivePrefix = {arXiv},
       eprint = {astro-ph/9701231},
 primaryClass = {astro-ph},
       adsurl = {https://ui.adsabs.harvard.edu/abs/1997PhRvL..78.2292W}
}

@ARTICLE{LiuWang13,
       author = {{Liu}, Ruo-Yu and {Wang}, Xiang-Yu},
        title = "{Diffuse PeV Neutrinos from Gamma-Ray Bursts}",
      journal = {\apj},
         year = 2013,
        month = apr,
       volume = {766},
       number = {2},
          eid = {73},
        pages = {73},
          doi = {10.1088/0004-637X/766/2/73},
archivePrefix = {arXiv},
       eprint = {1212.1260},
 primaryClass = {astro-ph.HE},
       adsurl = {https://ui.adsabs.harvard.edu/abs/2013ApJ...766...73L}
}

@ARTICLE{MuraseIoka,
       author = {{Murase}, Kohta and {Ioka}, Kunihito},
        title = "{TeV-PeV Neutrinos from Low-Power Gamma-Ray Burst Jets inside Stars}",
      journal = {\prl},
         year = 2013,
        month = sep,
       volume = {111},
       number = {12},
          eid = {121102},
        pages = {121102},
          doi = {10.1103/PhysRevLett.111.121102},
archivePrefix = {arXiv},
       eprint = {1306.2274},
 primaryClass = {astro-ph.HE},
       adsurl = {https://ui.adsabs.harvard.edu/abs/2013PhRvL.111l1102M}
}

@ARTICLE{Tamborra15,
       author = {{Tamborra}, Irene and {Ando}, Shin'ichiro},
        title = "{Inspecting the supernova-gamma-ray-burst connection with high-energy neutrinos}",
      journal = {\prd},
         year = 2016,
        month = mar,
       volume = {93},
       number = {5},
          eid = {053010},
        pages = {053010},
          doi = {10.1103/PhysRevD.93.053010},
archivePrefix = {arXiv},
       eprint = {1512.01559},
 primaryClass = {astro-ph.HE},
       adsurl = {https://ui.adsabs.harvard.edu/abs/2016PhRvD..93e3010T}
}

@ARTICLE{Kashiyama13,
       author = {{Kashiyama}, Kazumi and {Murase}, Kohta and {Horiuchi}, Shunsaku and {Gao}, Shan and {M{\'e}sz{\'a}ros}, Peter},
        title = "{High-energy Neutrino and Gamma-Ray Transients from Trans-relativistic Supernova Shock Breakouts}",
      journal = {\apjl},
         year = 2013,
        month = may,
       volume = {769},
       number = {1},
          eid = {L6},
        pages = {L6},
          doi = {10.1088/2041-8205/769/1/L6},
archivePrefix = {arXiv},
       eprint = {1210.8147},
 primaryClass = {astro-ph.HE},
       adsurl = {https://ui.adsabs.harvard.edu/abs/2013ApJ...769L...6K}
}

@ARTICLE{Zhang18,
       author = {{Zhang}, B. Theodore and {Murase}, Kohta and {Kimura}, Shigeo S. and {Horiuchi}, Shunsaku and {M{\'e}sz{\'a}ros}, Peter},
        title = "{Low-luminosity gamma-ray bursts as the sources of ultrahigh-energy cosmic ray nuclei}",
      journal = {\prd},
         year = 2018,
        month = apr,
       volume = {97},
       number = {8},
          eid = {083010},
        pages = {083010},
          doi = {10.1103/PhysRevD.97.083010},
archivePrefix = {arXiv},
       eprint = {1712.09984},
 primaryClass = {astro-ph.HE},
       adsurl = {https://ui.adsabs.harvard.edu/abs/2018PhRvD..97h3010Z}
}

@ARTICLE{Bonc19,
       author = {{Boncioli}, Denise and {Biehl}, Daniel and {Winter}, Walter},
        title = "{On the Common Origin of Cosmic Rays across the Ankle and Diffuse Neutrinos at the Highest Energies from Low-luminosity Gamma-Ray Bursts}",
      journal = {\apj},
         year = 2019,
        month = feb,
       volume = {872},
       number = {1},
          eid = {110},
        pages = {110},
          doi = {10.3847/1538-4357/aafda7},
archivePrefix = {arXiv},
       eprint = {1808.07481},
 primaryClass = {astro-ph.HE},
       adsurl = {https://ui.adsabs.harvard.edu/abs/2019ApJ...872..110B}
}

@ARTICLE{Devaki,
       author = {{Devaki Meduri}, Sri and {Desai}, Shantanu},
        title = "{Limit on high energy neutrino emission from Abell 119 using IceCube 10-year muon track data}",
      journal = {arXiv e-prints},
         year = 2026,
        month = may,
          eid = {arXiv:2605.11966},
        pages = {arXiv:2605.11966},
          doi = {10.48550/arXiv.2605.11966},
archivePrefix = {arXiv},
       eprint = {2605.11966},
 primaryClass = {astro-ph.HE},
       adsurl = {https://ui.adsabs.harvard.edu/abs/2026arXiv260511966D}
}

@ARTICLE{Qui24,
       author = {{Quirola-V{\'a}squez}, J. and {Bauer}, F.~E. and {Jonker}, P.~G. and {Brandt}, W.~N. and {Eappachen}, D. and {Levan}, A.~J. and {L{\'o}pez}, E. and {Luo}, B. and {Ravasio}, M.~E. and {Sun}, H. and {Xue}, Y.~Q. and {Yang}, G. and {Zheng}, X.~C.},
        title = "{Probing a magnetar origin for the population of extragalactic fast X-ray transients detected by Chandra}",
      journal = {\aap},
         year = 2024,
        month = mar,
       volume = {683},
          eid = {A243},
        pages = {A243},
          doi = {10.1051/0004-6361/202347629},
archivePrefix = {arXiv},
       eprint = {2401.01415},
 primaryClass = {astro-ph.HE},
       adsurl = {https://ui.adsabs.harvard.edu/abs/2024A&A...683A.243Q}
}

@article{Angu26,
    author = "Angulo-Valdez, Camila and others",
    title = "{EP260119a: A High-Redshift Gamma-Ray Quiet Fast X-ray Transient Probing a Potentially Hidden Population of Relativistic Explosions}",
    eprint = "2609.27127",
    archivePrefix = "arXiv",
    primaryClass = "astro-ph.HE",
    month = "9",
    year = "2026"
}

@ARTICLE{Bloom06,
       author = {{Woosley}, S.~E. and {Bloom}, J.~S.},
        title = "{The Supernova Gamma-Ray Burst Connection}",
      journal = {\araa},
         year = 2006,
        month = sep,
       volume = {44},
       number = {1},
        pages = {507-556},
          doi = {10.1146/annurev.astro.43.072103.150558},
archivePrefix = {arXiv},
       eprint = {astro-ph/0609142},
 primaryClass = {astro-ph},
       adsurl = {https://ui.adsabs.harvard.edu/abs/2006ARA&A..44..507W}
}

\end{document}